\documentstyle[12pt,epsf]{article}
\newcommand{\mo}{M{\o}ller}
\begin{document}
\LARGE \begin{center}
A high-precision polarimeter \\ 
\vspace*{1cm} \large
\vspace*{3mm} M. Hauger, A. Honegger,  J. Jourdan, G. Kubon, T. Petitjean, \\
D. Rohe,  I. Sick, G. Warren, H. W\"ohrle,  J. Zhao \\ 
Dept. f\"ur Physik und Astronomie, 
Universit\"at Basel, \\ CH-4056 Basel, Switzerland \\[3mm] 
R. Ent, J. Mitchell, \\
Jefferson Laboratory \\
Newport News, VA, 23606, USA\\[3mm]
D. Crabb, A. Tobias, M. Zeier, B. Zihlmann \\
Dept. of Physics, Univ. of Virginia \\
Charlottesville, VA, 22901, USA \\
\date{\today}
\end{center}
\vspace*{1.5cm}                 
\normalsize
\begin{center} Abstract \end{center}
\small
\begin{center}
\begin{minipage}[t]{1.0cm}
\hfill
\end{minipage}
\begin{minipage}[t]{12cm}
We have built a polarimeter in order to measure the electron beam polarization
in hall C at JLAB. Using a  superconducting solenoid to drive the 
pure-iron target foil into 
saturation, and a symmetrical setup to detect the \mo\ electrons in coincidence,
 we achieve an accuracy of $<$1\%. This sets a new standard for 
\mo\ polarimeters.
\end{minipage} \\
\end{center}
 \normalsize
\hspace*{1cm} PACS: 29.25.Pj, 29.27.Hj, 34.80.Nz, 33.55.Fi
\newpage

\section{Introduction}
Today, electron scattering experiments often involve the use of high energy 
(GeV) polarized 
electrons. Recent examples are measurements of the spin structure of the
nucleon via $\vec{e}-\vec{^1 H}$ and $\vec{e}-\vec{^2 H}$ deep inelastic
scattering\cite{Abe95,Abe96}, or measurements of the charge form factor of the neutron via
neutron knockout experiments ($\vec{e},e'n)$ on $\vec{^3 He}$ or 
$\vec{^2 H}$ \cite{Day89b,Rohe98}. Such experiments are becoming 
increasingly more practical as electron sources providing beams of high intensity and
high polarization \cite{Alley95} become standard equipment at the various 
electron accelerator laboratories.

For these experiments with polarized electrons, one needs polarimeters 
to measure the  polarization of the high-energy beam. Standard methods are 
based on Compton 
($\vec{e}-\vec{\gamma}$) or \mo\
($\vec{e}-\vec{e}$ ) scattering. For these processes the analyzing power can
be accurately calculated, so a measurement 
of the polarization asymmetry of the cross section yields the beam 
polarization once the polarization of the photon or the $\vec{e}$-target
is known. Compton scattering in general leads to rather low rates, thus 
Compton polarimeters are preferably used in connection with large electron
beam intensities (storage rings)
\cite{Gustavson79,Barber93}. 
\mo\   scattering allows for much larger rates, and therefore can also 
be used for cases where the beam intensity is low, such as required in connection 
with cryogenic polarized targets which typically are limited to beam currents 
$<$100nA. The polarimeter we describe in this paper belongs to the \mo\
 scattering category\cite{Wagner90b,Arrington92,Steiner98}, and has been built 
in connection with experiments using polarized cryogenic $N\vec{^1 H}_3$ and 
$N\vec{^2 H}_3$ targets at JLAB.

\mo\ polarimeters are based on $\vec{e} + \vec{e} \rightarrow e + 
e$ scattering.  Since this is a pure QED process, its cross-section 
can be calculated to very high precision.  For longitudinally ($||$) 
polarized beam (polarization $P_{b}^{||}$ parallel to the z-axis) and target 
($P_{t}^{||}$) electrons, the 
cross-section is expressed in the centre of mass (c.m.) frame as:
\begin{equation}
\label{pol_xsect}
\frac{d \sigma}{d \Omega}= \frac{d \sigma_{\circ}}{d \Omega} \left[ 1 +
P_{t}^{||}P_{b}^{||}A_{zz}(\theta) \right]
\end{equation}
where the unpolarized cross-section is given by 
 $d \sigma_{\circ}/d \Omega =
(\alpha(4-\sin^{2}\theta)/2m_{e} \gamma \sin^{2}\theta)^{2}$ at high 
energy, and the
analyzing power by $A_{zz}(\theta)=
-\sin^{2}\theta(8-\sin^{2}\theta)/(4-\sin^{2}\theta)^{2}$. One can
effectively measure the beam polarization by comparing the cross-section 
asymmetry for beam and target spins aligned parallel and anti-parallel: 
\begin{equation}
\label{asym}
\epsilon=\frac{\frac{d \sigma^{\uparrow \uparrow}}{d \Omega} -
\frac{d \sigma^{\uparrow \downarrow}}{d \Omega}}{\frac{d \sigma^{\uparrow 
\uparrow}}{d \Omega} + \frac{d \sigma^{\uparrow \downarrow}}{d \Omega}}=
A_{zz}(\theta)P_{b}^{||}P_{t}^{||}.
\end{equation}
At 90 degrees (c.m.), the analyzing power is large, $A_{zz}=-7/9$, 
and so is the cross-section, $d \sigma_{\circ}/d 
\Omega_{lab}$=17.9 fm$^2$ sr$^{-1}$. 
Further, these quantities are energy independent for large $\gamma$ 
$=E/m_{e}c^{2}$. 

The expressions used in eqn.~\ref{pol_xsect} follow from the lowest 
order diagrams of this process.  A calculation has been performed for 
all diagrams to order $\alpha^{4}$~\cite{DeRaad75}, and although a 
significant impact is found for $d \sigma_{\circ}/d \Omega$, a 
negligible effect is found for $A_{zz}$.  

Despite the large analyzing power of --7/9, \mo\ scattering is not easy to exploit. In
ferromagnetic $\vec{e}$-targets, only 2 of the typically 25 electrons are 
polarized, leading to an effective target polarization of only $\sim$8\%. 
The reliable determination of the beam polarization requires tight 
control over systematic errors, and  high statistics.

In order to achieve a precision measurement of the beam polarization, certain
limitations have to be overcome. In earlier polarimeter designs
\cite{Wagner90b,Arrington92}, three factors
contributed significantly to the overall error: achievable statistics,
uncertainties due to background contributions (mainly Mott scattering), and 
the error in the
determination of the target electron polarization $P_t$. Up till recently, 
no system had obtained an uncertainty better than 3\%~\cite{Feltham95} and many 
were much worse. Furthermore, almost all work neglected the influence of the 
atomic
motion of the target electrons on the effective analyzing powers, 
an effect only recently identified by Levchuk~\cite{Levchuk94}. 

In the present polarimeter we reduce the background contribution uncertainty
by using a coincidence detection system; this allows us to eliminate the 
dominating background from Mott scattering which does not lead to coincident
$e-e$ pairs. To enhance the statistical 
precision without requiring long data acquisition time, we employ a 
 large acceptance for detection of the scattered electrons. Using a large 
acceptance at the same time
allows us to reduce the influence of the atomic motion of the electrons in the 
\mo\ target on the effective analyzing powers A$_{zz}$, to the point where
it easily can be corrected for. 
In order to eliminate the then dominant error from the uncertainty in 
the polarization $P_t$ of the electrons in the magnetized 
ferromagnetic foil we use a novel approach \cite{Robinson94,Bever97} which employs
foils made from {\em pure} iron, magnetized out of plane to saturation 
using a 4T field. 
For pure iron in saturation, the electron polarization is known with 
great accuracy.

In the present paper, we describe the polarimeter built and used in hall C
at JLAB. We show that with the approach taken we are able to reduce the
uncertainty to below the 1\%-level.

\section{Set-up}
\begin{figure}[htb]
\vspace*{1cm}
\centerline{\mbox{\epsfysize=6cm\epsffile{
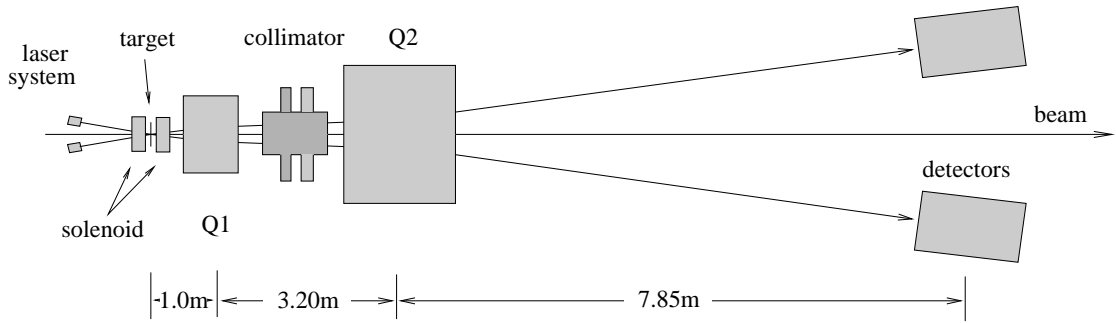
}}}
\vspace*{0.5cm}
\caption{
Layout of the hall-C polarimeter. 
 \label{setup}}
\vspace*{0.5cm}
\end{figure}

The general layout of the polarimeter is shown in fig.\ref{setup}.
The incoming electron beam of energy of (typically) 1 -- 6 GeV is focused onto the 
\mo\ target. This target is made from a thin foil of pure iron oriented 
perpendicular to the electron beam, and which is magnetized using a superconducting 
solenoid producing a 4T field in the direction of the beam. The scattered 
electron and the recoiling target electron, which emerge in the horizontal 
plane at angles between
1.83 and 0.75 degrees in the laboratory, are focused using a first quadrupole 
Q1. The desired scattering angles are selected by a set of collimators.
The electrons then are defocused using the quadrupole Q2, and detected in 
coincidence using two symmetrically placed hodoscope counters and lead-glass 
detectors. We below describe the individual elements in more detail.

\underline{Target} 

The guiding principle for the choice of the target has been investigated by 
deBever {\em et al} \cite{Bever97}. As a source of polarized target electrons 
we use {\em pure} 
iron. The polarizable electrons, two of the 25 per atom, are polarized 
using a 4T field in beam direction, perpendicular to the iron foil. This setup
differs in a number of ways from the standard one involving  foils of an iron 
alloy, oriented at 
$\sim$20$^\circ$ relative to the beam and polarized in-plane using a magnetic field of
the order of 0.01T. 
\begin{itemize} 
\item
The spin polarization for pure iron in saturation is known with excellent 
precision \cite{Reck69}. This precision basically comes from the fact that, in 
saturation, the properties of an isolated iron atom and the atoms in a foil 
are the same. In this case the spin polarization can be measured using 
large pieces of iron;  measurements of the magnetization together with measurements 
exploiting the 
Einstein-deHaas effect allow to separate the spin contribution from the orbital
contribution to the magnetization, the latter being  the main source 
of uncertainty for alloys. The spin polarization for pure iron has been measured to
about 1/4\% accuracy \cite{Reck69}. 
\item Saturating a pure iron foil {\em out of plane} comes at the expense 
of requiring a high magnetic field, $\sim$4T. Such fields today can be produced
easily using small superconducting split coils, and the longitudinal B-field 
has little effect on the incoming and scattered electrons.
\item Since the foil is saturated {\em brute-force} no delicate absolute measurements
 of the foil magnetization (polarization) using in-situ pickup coils are 
required. 
\item Under conditions of high beam current, leading to heating of the iron 
target, the  decrease in foil polarization can easily be measured using a 
Kerr apparatus \cite{Bever97}. (This Kerr apparatus has been built and tested,
but was not employed in the first experiment described here as the currents 
used were very low).
\item The lack of need to measure the foil magnetization allows us to use a 
large dynamical range of foil thicknesses, as governed by the beam intensities 
used by the main experiment. At the same time, rotation of the foil to 
spread the heat could easily be performed if much larger beam intensities 
should be required.
\item Since the target is perpendicular to the beam, the usually needed 
corrections for the cosine of the $\sim 20^\circ$ angle between target 
polarization and 
beam direction are not needed; this also eliminates the  
measurements with symmetrical target orientation often required to 
eliminate uncertainties in the target angle when dealing with slightly
warped target foils.
\end{itemize}

The \mo\ target is placed on a target ladder which, by remote control, can 
be moved horizontally into the beam. This target ladder carries foils of 
several different thicknesses,
together with a viewing screen to check the alignment of the beam.

\underline{Solenoid}

The superconducting split-coil solenoid used to polarize the iron target was purchased 
from Oxford, and is run using an IPS120-10 power supply. The split coils 
produce a maximal field of 4T. The coils have a  6.7cm diameter
bore for the beam, and a sideways access of 3.5cm width and 7.6cm height for
the target ladder.  
The liquid nitrogen and helium required for operation of the solenoid is 
supplied by the hall-C cryosystem.

\underline{Magneto-Optical System}

We use a two-quadrupole system to increase the angles of the scattered- and recoil
electrons such as to be able to place the detectors at a reasonable distance
from the beamline. The quadrupoles deflect the \mo\ electrons, without affecting
the incident beam which goes through the center of Q1, Q2 and is virtually
undisturbed given its low emittance. Besides the increase of the separation 
between \mo\ electrons and beam, Q2 also provides, in combination with the 
collimator, an energy analysis of the \mo\ electrons. This greatly reduces 
backgrounds.

Studies of various layouts have shown 
that a system with {\em two} quadrupoles is much more flexible than the usual 
one-quadrupole or magnet plus septum systems; for the entire energy range of 
interest, 1 -- 6 GeV, 
the \mo\ electrons can be imaged onto the detectors without any change in 
geometry. The main function of Q1 occurs at low incident electron energy.
There, the \mo\ electrons for 90$^\circ$ CM-angle are produced at 
comparatively large angles in the laboratory system, and Q1 focuses these  
electrons into the acceptance of Q2. At large electron energy, Q1 has 
virtually no effect. With the two-quadrupole setup, Q2 can be placed at a 
larger distance from the target, thus maximizing the overall deflection of 
the \mo\ electrons. 

\begin{figure}[htb]
\vspace*{1cm}
\centerline{\mbox{\epsfysize=6cm\epsffile{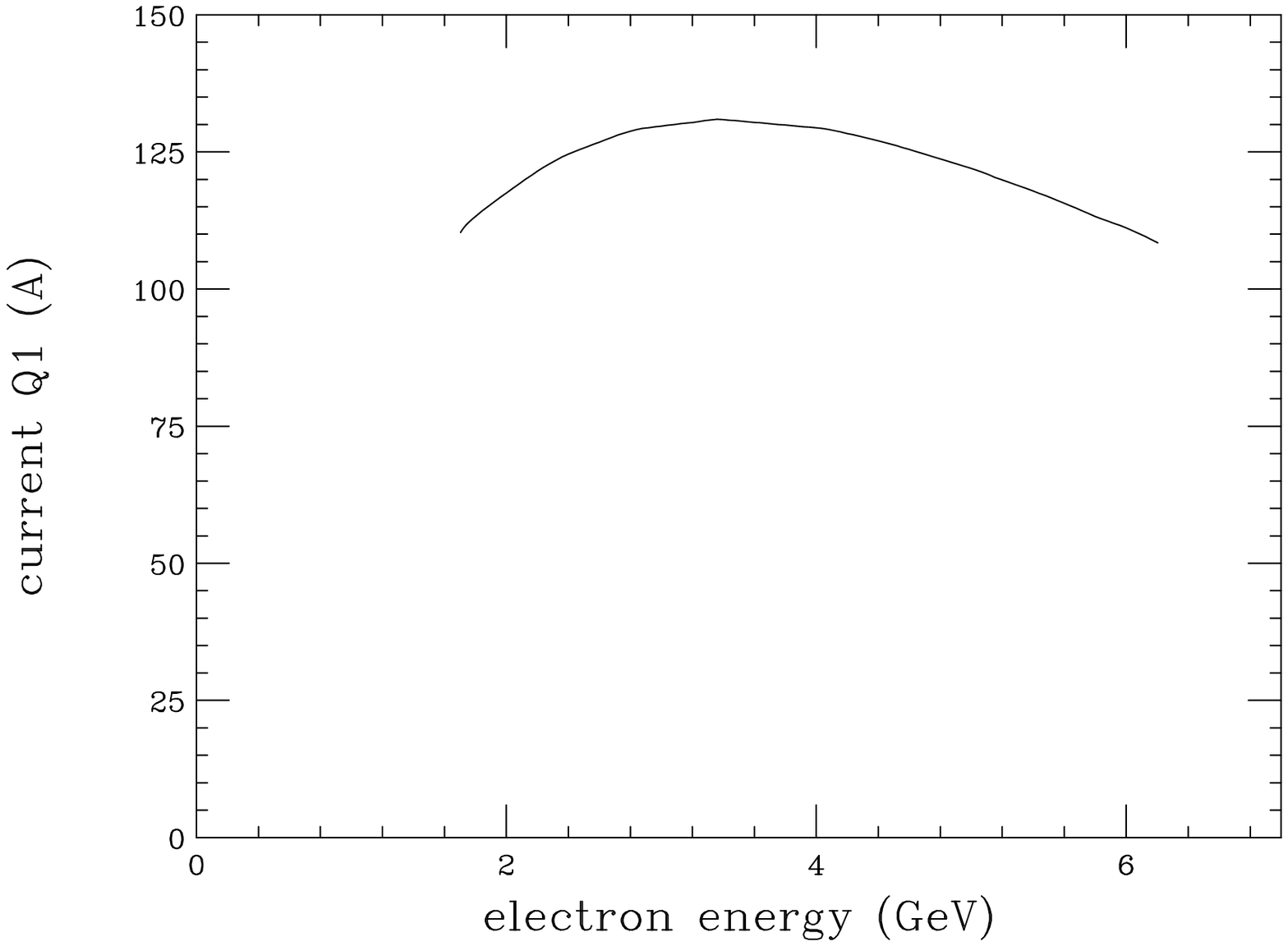}\epsfysize=6cm\epsffile{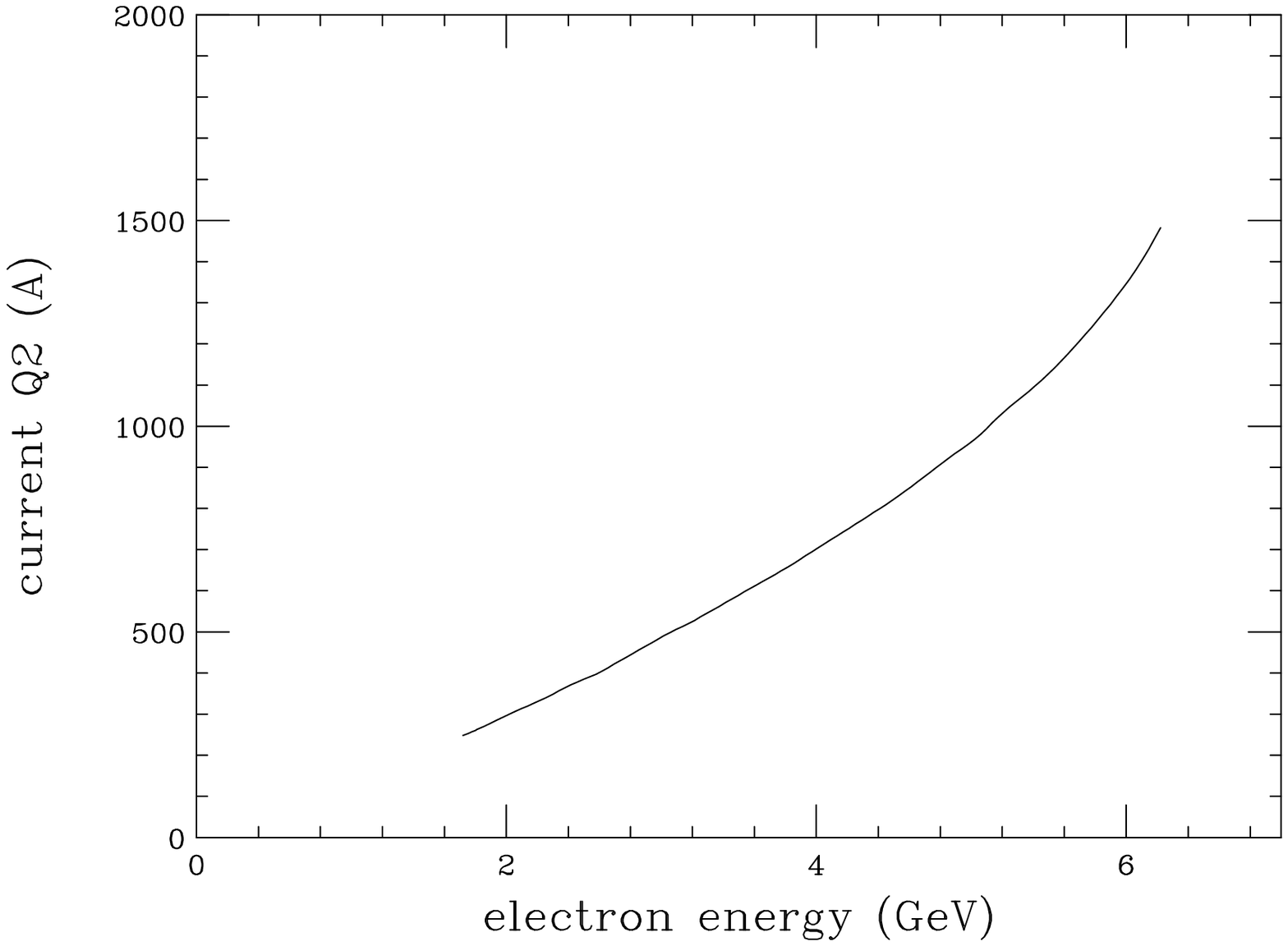}
}}
\vspace*{0.5cm}
\caption{
Setting of Q1, Q2 for optimal imaging of the \mo\ electrons onto the detectors.
 \label{tune}}
\vspace*{0.5cm}
\end{figure}

This two-quadrupole system allows us to maintain, over the entire energy range
of 1 -- 6 GeV, 
an image of the 90$^\circ$ electrons of elliptical shape with an axis-ratio of
$<$2 at the location of the detectors. With such an image, a clean
selection of coincident electrons using slits in front of the detectors, and
a useful measurement of the scattering angles using the hodoscope is
possible.
The optimal tuning of Q1, Q2, calculated using a MC simulation of the setup, 
is given in fig.\ref{tune}.

\underline{Adjustable collimators}

\mo\ scattering produces a spectrum of scattering angles; for polarization measurement,
angles around 90$^\circ$ in the CM system are of particular interest given
the maximal analyzing power of 7/9. At the same time, Mott scattering (from
nuclei) produces a large flux of scattered electrons at small scattering angle,
which one would like to suppress as far as possible.
\begin{figure}[htb]
\vspace*{1cm}
\centerline{\mbox{\epsfysize=7cm\epsffile{
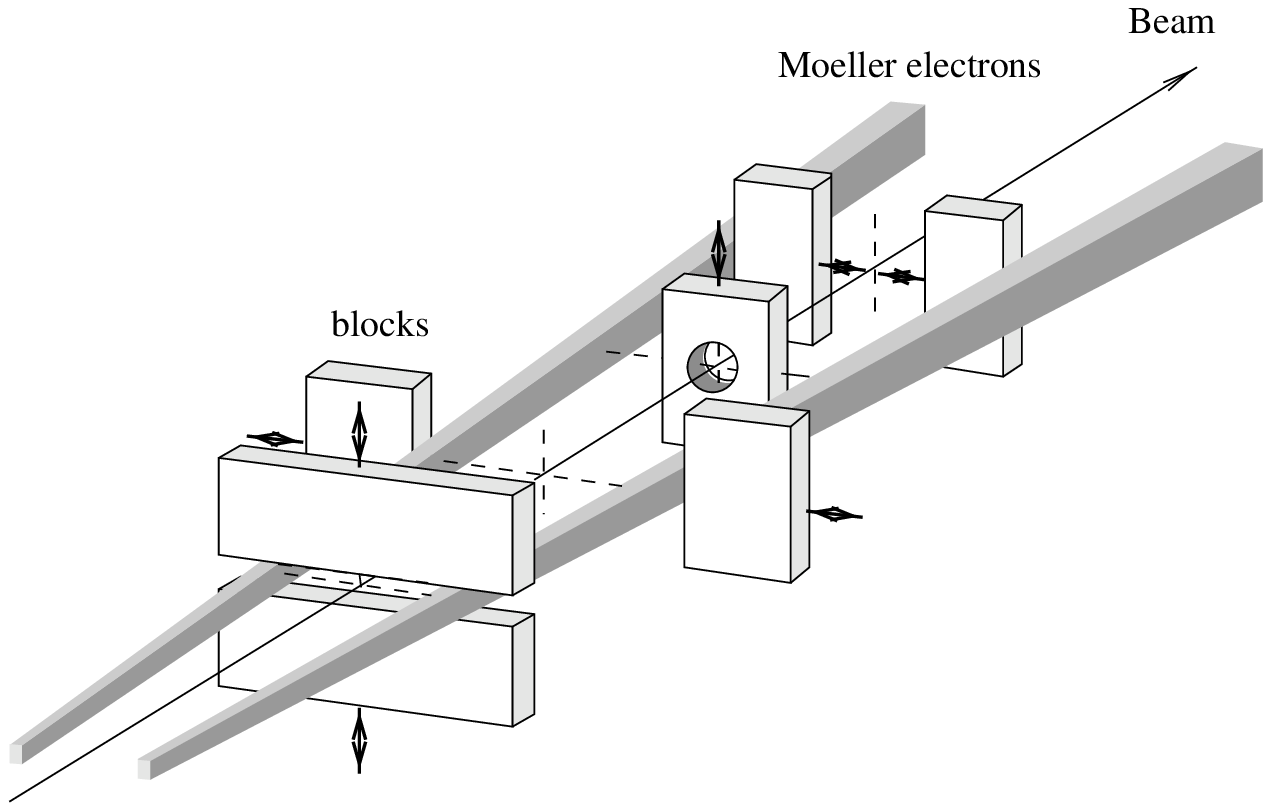
}}}
\vspace*{0.5cm}
\caption{
collimator system used, showing the 6 moveable jaws, together with the 
block covering the central part.
 \label{coll}}
\vspace*{0.5cm}
\end{figure}

The collimator system has been designed to select a range of scattering angles, and 
to cut off electrons at both smaller and large angles. This is achieved by a 
set of six moveable jaws (see figure \ref{coll}). A seventh collimator with 
fixed acceptance in the center eliminates the electrons that could pass on 
the small-angle side of the inner horizontal collimator. When the polarimeter 
is not in use, these collimators are all removed by remote control.

The collimator jaws are made from densimet, $\sim$  8cm thick (22 radiation 
lengths). With this thickness, all unwanted electrons are removed, or
loose so much energy that they can no longer give large enough a signal in the
lead-glass total absorption counters.

The selection of scattering angles by the collimators is only a rough one, and
is made such as to be less constraining than the selection made by the 
slits  in front of the hodoscope. The main function of the collimators 
then is to 
stop electrons which otherwise could hit the vacuum enclosure and get, through 
uncontrolled pathways, to the detectors. 

The collimators are  placed {\em before} Q2, such that the energy-analysis performed 
by Q2 removes eventual low-energy electrons that are produced in the jaws.    

\underline{Slits}

In front of the detector package, two slits define the actual angular 
acceptance of the polarimeter. These slits are about 12cm wide in 
horizontal direction,  and have a tapered opening of $\pm$2cm -- $\pm$3cm in the
vertical direction, such as to select a constant bin in out-of-plane angle 
$\phi$. One of the slits has a somewhat larger acceptance, in order to
insure that the other slit is the one that sets the angular acceptance.
Simulations have shown that this arrangement minimizes the Levchuk effect 
(see below).

The slits are made from lead, and are 9 radiation lengths thick.

\underline{Detector package} 

The main detectors identifying the \mo\ electrons are the lead-glass total 
absorption counters. The blocks have dimension of 20x14x23cm$^3$
in order to contain the entire shower produced by the \mo\ electrons, and
are made from SF2 lead glass. 
Measurement of the amplitude of the light signal, observed with one 5'' photo
tube for each block, provides a discriminating signal for the presence of an 
electron of the appropriate energy. Measurement of two such electrons
in coincidence provides a virtually background-free identification of
\mo\ electron pairs.
\begin{figure}[htb]
\vspace*{1cm}
\centerline{\mbox{\epsfysize=5cm\epsffile{
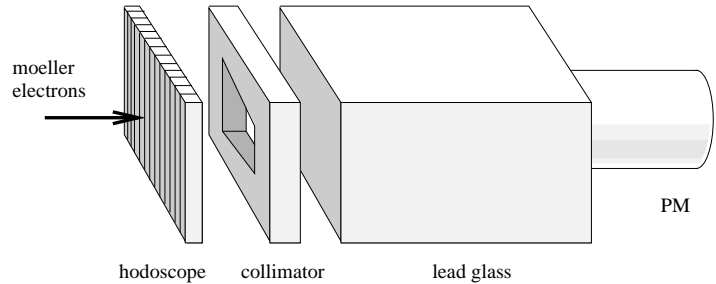
}}}
\vspace*{0.5cm}
\caption{
Package of hodoscope, slits and lead glass total absorption counter.
 \label{det}}
\vspace*{0.5cm}
\end{figure}

For the actual measurement of the beam polarization, only the coincidence rate 
of the total absorption counters is used. The number of coincidences can be 
measured over a very large dynamical range of the rate with very little dead 
time, and minimal investment in terms of electronics.

In front of the lead-glass total absorption counter, two hodoscopes provide 
information on the location of the coincident electrons. This information is 
used during the setup of the polarimeter mainly, it need not be recorded during 
the actual measurement of the polarization.

The hodoscopes have 14 horizontal channels each. Each channel consists of a 
bar  of scintillator, 8x12mm$^2$ in cross section and 80mm high, viewed by a 
8mm diameter photomultiplier. Two of the channels, number 3 and 12, are 
vertically split into 
two halves, such as to also get information on the vertical tuning. The information
from the hodoscopes is read out only once the pair of lead-glass detectors 
has identified a coincident \mo\ pair.  

In front of the hodoscopes, we have placed a layer of lead, 1cm thick. This
serves to remove low-energy background which could lead to high count rates
in the hodoscope, and it leads to the development of a shower which causes an
increased energy deposit in the hodoscope. 

\underline{Electronics}

The main information provided by the polarimeter is the  rate 
of the two lead-glass detectors measured using  two photo multipliers, 
fast *5 amplifiers, clipping of the analog signals  and fast 
discriminators (5 and 10ns pulse width). Prompt and delayed Left-Right (L-R)
coincidences are  registered
 during a run using VME scalers. The scalers are read out upon completion 
of the run or a flip of the electron polarization at the source.  

The analog  signals from the hodoscopes  are sent through ECLINE discriminators
and the pattern of both hodoscopes is registered  for valid L-R coincidences
using  fast CAMAC memories (LeCroy 4302) capable of storing 16k events. 
During data taking runs, the hodoscope information is recorded  for a small 
fraction of the L-R coincidences only, as it 
only  serves as a check of the proper alignment.   

\underline{Control}

The entire \mo\ system including the cryogenics needed for the solenoid
is controlled from two Graphics User Interfaces (GUI). The GUIs use the 
Experimental Physics and Industrial Control System (EPICS), a 
set of software tools and applications  jointly 
developed by Los Alamos National Laboratory  and
Argonne National Laboratory \cite{Dalesio91}. The control programs communicating
with the hardware over various interfaces running
on front-end computers in VME crates. The interface to the backend
user program or GUIs is the EPICS data base.

\section{Results}
We have carried out a number of measurements of the beam polarization, 
accompanied by a number of sensitivity tests. Some of the results are described
below.

For the measurements carried out, we have used beam currents of typically 
1$\mu$A;
the main experiment ran  usually at 100nA. In tests the polarimeter has been 
successfully used at currents up to 8$\mu$A. As \mo\ target, we used  pure iron
foils, 4 and 10 $\mu$m thick.

Tuneup for the polarization measurements involved focussing of the beam onto
the polarimeter target, which was located some 30m upstream of the spectrometers
used for the main experiment.
 Using the viewers on the polarimeter target ladder, and another viewer
placed 12m upstream, the position and direction of the incoming electron beam 
were adjusted, to $\pm$0.5mm. For some of the actual measurements, the fast 
rastering of the electron was then turned on, leading to a beam distributed over 
$\pm$0.5 or $\pm$1mm. In order to keep the beam centered downstream of the 
polarimeter after ramping up the fields of the solenoid and Q1, Q2,
only minor adjustments of steering coils downstream of the polarimeter were 
needed.
\begin{figure}[htb]
\centerline{\mbox{\epsfysize=7cm\epsffile{
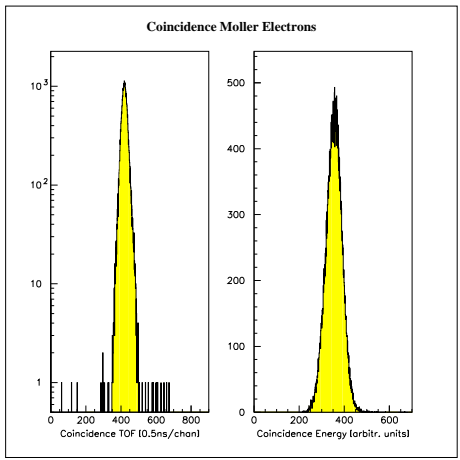
}}}
\caption{
Time- and energy spectrum of the \mo\ electrons. For the energy distribution 
the sum of the two lead-glass detector analog signals, which should correspond 
to the incident electron energy, is shown.  
 \label{tof}}
\end{figure}

The beam current was measured using RF cavities with high bandwidth
and signal to noise ratio. The cavities are temperature stabilized
to 0.3$^\circ$. The 1.497 GHz signal from the cavity 
is down-converted to 100kHz and subsequently 
converted into a DC-voltage which is proportional to the beam current. 
The DC Voltage goes to a 1MHz voltage to frequency converter, the output of 
which is integrated using  a VME scaler. An absolute 
calibration of these cavities is obtained using an independent Unser
monitor \cite{Unser81}. In several calibration periods two-minute runs with
alternating periods of beam-on and beam-off are made at various beam
currents. The beam off
periods allow to measure the zero offset of the cavity and  Unser monitor, 
the comparison of the  Unser monitor and  the cavities during beam-on yields
the gains of the cavities. 

During the data reduction, we have also analyzed the asymmetry of the 
charge accumulated for the two polarization states. This asymmetry was usually 
below 5 10$^{-4}$. This leads to negligible errors in the polarization measurements 
due to eventual non-linearities of the current monitors, or zero-offsets.
The polarity of the electron polarization was flipped once a second at the 
polarized source.
\begin{figure}[htb]
\centerline{\mbox{\epsfysize=7cm\epsffile{
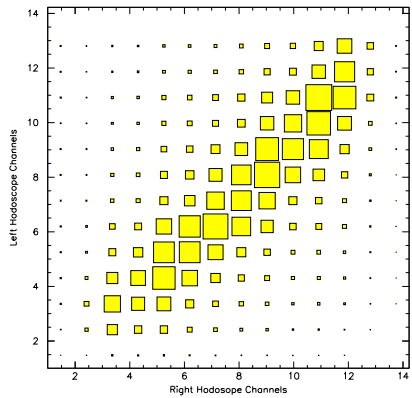
}}}
\caption{
Distribution of \mo\ events on the hodoscopes. 
 \label{hodo2d}}
\end{figure}

As pointed out above, the main observable used for the determination of the
beam polarization is the coincidence rate between the two lead-glass 
detectors. Fig.\ref{tof} shows the spectrum of the time difference between 
the signals in the two detectors registered in sampling-mode in parallel.  
Under the conditions used for these measurements, the rate of accidental 
coincidences is obviously very small. The accidentals have been measured via a 
delayed coincidence, and subtracted. 

In fig. \ref{hodo2d} we show the distribution of events on the hodoscopes. 
The coincidence signals of interest populate the region of the diagonal, as
expected for \mo\ scattering. The higher density of events towards the upper
left-hand side reflects electrons that have lost energy due to radiative processes.

\begin{figure}[htb]
\centerline{\mbox{\epsfysize=7cm\epsffile{
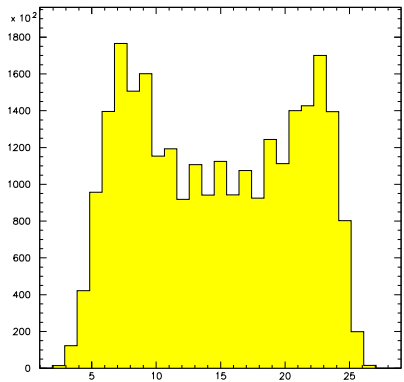
}}}
\caption{
Projection of the hodoscope events onto the diagonal of fig.
\protect{\ref{hodo2d}}.
The minimum at channel 15 corresponds to the minimum at 90$^\circ$ CM scattering
angle of the \mo\ cross section.
 \label{proj}}
\end{figure}

Improper tuning of Q2 can lead to a  shift of the ridge away from the diagonal,
a shift that is very easy to observe. The same is true for a mistune of Q1. 
A combined mistune of Q1 and Q2 leads to a shift along the diagonal. This
can easily be observed by looking at the projection of the events onto
the diagonal, see fig. \ref{proj}. Such a mistune leads to a shift of the 
minimum along the diagonal. 

\begin{figure}[htb]
\centerline{\mbox{\epsfysize=6cm\epsffile{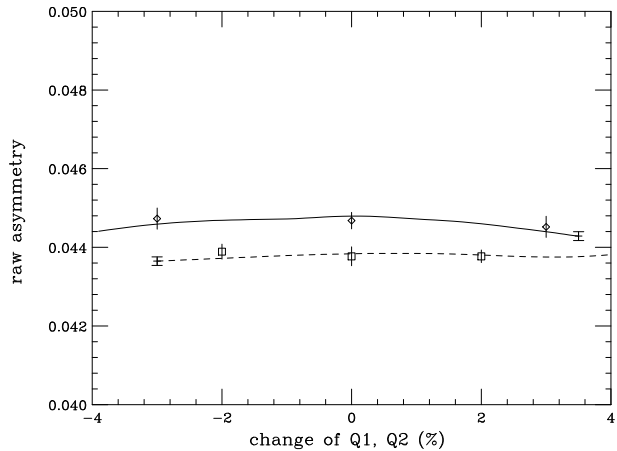}}}
\caption{
Change of the raw asymmetry with detuning of the quadrupoles Q1 (solid) and
 Q2 (dashed). The curves represent the MC simulation, the bar indicates
the statistical uncertainty of the simulation. 
 \label{moll3}}
\vspace*{0.5cm}
\end{figure}

We have investigated the effect of a mistune of Q1, Q2 both via our Monte-Carlo
simulation, and experimentally. In fig. \ref{moll3} we show the change of the 
analyzing power observed when detuning the current of Q1 (Q2) by 2\%(1\%).
The observed sensitivity to the tuning of Q1, Q2 corresponds to the one expected
from the simulation. 

The time needed to acquire the statistics for precise measurements of the beam
polarization is quite short. At $\sim$70\% beam polarization as available during 
the runs used here it took 5 minutes to acquire the  9$\cdot$10$^6$ events needed 
to get 
a 1\% relative uncertainty on the beam polarization. If needed, this time could be
further shortened by using  iron foils thicker than 10$\mu$m.

We have found that tuning of the polarimeter is very straightforward. 
Potential errors in the set-up are detected easily, and corrections to be
applied to the raw asymmetry (accidentals, dead time, ...) are very small.  

\section{Sensitivity studies}
The uncertainty of the polarization measurement is determined by the knowledge
of the foil polarization, the statistical error on the count rate asymmetry,
and various errors in the set-up and tuning of the hardware. 

We have described in the introduction the basic principle of our polarimeter
which was chosen such as to minimize the usually dominating error of the 
spin polarization of the atomic electrons. For pure iron, the spin polarization
in saturation is known to 0.25\%. 

The uncertainties due to the various potential mis-tuning of different elements
have been minimized in the design, by selecting a set-up which is left-right
symmetrical, and exploits the \mo\ angular distribution around the symmetry
point at 90$^\circ$ CM angle.  As a consequence, all errors from geometry 
and tuning drop
out in first order; only {\em quadratic terms} come in. This makes the 
polarimeter very insensitive to the various errors.

In order to explore the sensitivity, we have used the Monte-Carlo simulation 
program. In table \ref{sens} we quote for the various potential sources of 
error due to 
alignment of the incoming beam, 
alignment of the quadrupoles,
mistune of the quadrupoles and
uncertainty in the angle-acceptance defining slits.
We also quote the effect of the uncertainty due to imperfect corrections for 
the Levchuk effect (motion of the atomic electrons) and multiple scattering of
electrons in the foil.

\begin{table}[htb] 
\begin{center}
\begin{tabular}{llr}
\hline
source & uncertainty & effect A \\
\hline
beam position x & 0.5mm & 0.15\% \\
beam position y & 0.5mm & 0.03\% \\
beam direction x & 0.15mr & 0.04\% \\
beam direction y & 0.15mr & 0.04\% \\
current Q1 & 2\% & 0.10\% \\
current Q2 & 1\% & 0.07\% \\
position Q2 & 1mm & 0.02\% \\
multiple scattering & 10\% & 0.12\% \\
Levchuk effect & 10\% & 0.30\% \\
position collimator & 0.5mm &  0.06\% \\
target temperature  & 50\% & 0.05\% \\
direction B-field & 2$^\circ$ & 0.06\% \\
value B-field & 5\% & 0.03\% \\
spin polarization in Fe & & 0.25\% \\
\hline
total & & 0.47\% \\
\hline
\end{tabular}
\caption{\label{sens} Sensitivity of effective analyzing power to various
sources of uncertainties.}
\end{center}
\end{table} 

For the alignment of the beam, we quote the effect of uncertainty due to
the  beam position measurement. For the placement of the 
elements such as quadrupoles and collimators, we use an uncertainty which is
much larger than the uncertainty quoted by the surveyors. For the uncertainty
in the focal strength of Q1, Q2 we use as values the deviations that can 
easily be recognized when looking at figs. \ref{hodo2d},\ref{proj}. 

For the Levchuk effect, we use 10\% of the change in the analyzing power obtained
when turning off the Levchuk effect entirely. The Levchuk effect has been calculated
using the program of \cite{Swartz95} adapted to our conditions; it amounts to 
3.03\% at 4GeV, and similar values over the entire energy range. 
The Levchuk effect in our case leads to a small change of the analyzing power 
 as a consequence of the large acceptance of the 
polarimeter, and the fact that we use a slightly asymmetric collimator for the two
sides. We give the calculated value a 10\% uncertainty in order to account for 
potential uncertainties in the atomic wave functions employed in the 
calculation.

For the multiple scattering effect, we quote an uncertainty of 10\% of the
multiple scattering effect, in order to account for possible uncertainties
in the Fe-foil thickness. 

We also include an uncertainty for the local temperature of the Fe foil due
to beam heating. This uncertainty is based on an assumed 50\% uncertainty in 
the beam spot size. During test using rastering of the beam we have verified at 
higher beam currents that the effect of target heating on the foil polarization
is indeed negligible.

 At beam intensities much higher than the 
1$\mu$A used here, the uncertainty in the temperature could become a substantial 
source of error. It can be eliminated by using thinner foils (yielding
better radiation cooling), and
by putting into operation the Kerr apparatus we have developed for measuring
the change of foil polarization occurring due to the electron beam
\cite{Bever97}.   

Table \ref{sens} shows that the cumulated uncertainty in the beam polarization
measurement is very small.

\section{Conclusion}
We have described in this paper a polarimeter designed to measure the 
polarization of the JLAB electron beam at energies between 1 and 6 GeV. 
This polarimeter exploits the  measurement of the cross section asymmetry in 
electron-electron scattering, the analyzing power of which is accurately known.
 The polarimeter is based on the idea, put 
forward in \cite{Bever97}, to use a pure iron foil in saturation at 4T field 
as a source of target electrons; for this system the target electron spin 
polarization has been very accurately measured.

The polarimeter designed and built for hall C detects the scattered and 
recoiling electron using total-absorption counters, and registers coincidence 
events only. This makes the setup very insensitive to all sorts of background
(Mott scattering in particular). The polarimeter has an entirely symmetric 
set-up; this eliminates all potential errors in first order. The quadratic
effect  of errors then becomes very small.

This polarimeter allows us to measure the electron beam polarization with 
$<$1\% statistical error in 5 minutes. The systematic error is $\sim$0.5\%.
This represents a major improvement in the accuracy of \mo\ polarimeters 
which up to now were limited to systematic uncertainties in the 3\% range. 

If needed, the accuracy of our polarimeter can, with little effort, be further
increased. Measurement of the target electron depolarization at high beam
intensities has been shown to work using the Kerr effect \cite{Bever97}.
The contribution of most other sources of systematic errors can be reduced in a 
straightforward fashion, leaving as the main contribution ultimately the 
knowledge of the electron spin polarization in iron ($\pm$0.25\%).


\begin{thebibliography}{10}

\bibitem{Abe95}
K.~Abe {\em et al.}
\newblock {\em Phys. Rev. Lett.}, 75:25, 1995.

\bibitem{Abe96}
K.~Abe {\em et al}.
\newblock {\em Phys. Rev. Lett.}, 76:587, 1996.

\bibitem{Day89b}
D.Day, M.~Farkondeh, K.~Giovanetti, J.~Lichtenstadt, R.~Lindgreen, J.S.
  McCarthy, R.Minehart, B.~Norum, D.~Pocencic, R.~Sealock, J.Jourdan,
  G.~Masson, and I.~Sick.
\newblock {\em CEBAF proposal 93-026}.

\bibitem{Rohe98}
D.~Rohe.
\newblock {\em Thesis, University of Mainz}, 1998.

\bibitem{Alley95}
R.~Alley {\em et al.}
\newblock {\em Nucl. Instr. Meth.}, A365:1, 1995.

\bibitem{Gustavson79}
D.~Gustavson, J.J. Murray, T.J. Phillips, R.~Schwitters, C.K. Sinclair, J.R.
  Johnson, R.~Prepost, and D.E. Wiser.
\newblock {\em Nucl. Instr. Meth.}, l65:177, 1979.

\bibitem{Barber93}
D.P.~Barber {\em et al.}
\newblock {\em Nucl. Instr. Meth.}, A329:79, 1993.

\bibitem{Wagner90b}
B.~Wagner, H.G. Andresen, K.H. Steffens, W.~Hartmann, W.~Heil, and E.~Reichert.
\newblock {\em Nucl. Instr. Meth. A}, 294:541, 1990.

\bibitem{Arrington92}
J.~Arrington, E.J. Beise, B.W. Filippone, T.G. O'Neill, W.R. Dodge, G.W.
  Dodson, K.A. Dow, and J.D. Zumbro.
\newblock {\em Nucl. Instr. Meth.}, A311:39, 1992.

\bibitem{Steiner98}
P.~Steiner, A.~Feltham, I.~Sick, and B.~Zihlmann.
\newblock {\em Nucl. Instr. Meth. A}, 419:105, 1998.

\bibitem{DeRaad75}
L.L. DeRaad and Y.~Ng.
\newblock {\em Phys. Rev.}, D11:1586, 1975.

\bibitem{Feltham95}
A.~Feltham and Ph. Steiner.
\newblock {\em Proc. Trieste Conf. Int. En. Phys.}, World Scient.:p47, 1995.

\bibitem{Levchuk94}
L.G. Levchuk.
\newblock {\em Nucl. Instr. and Meth.}, A345:496, 1994.

\bibitem{Robinson94}
S.~Robinson.
\newblock {\em Thesis, Basel}, 1994.

\bibitem{Bever97}
L.~deBever, J.~Jourdan, M.~Loppacher, S.~Robinson, I.~Sick, and J.~Zhao.
\newblock {\em Nucl. Instr. Meth. A}, 400:379, 1997.

\bibitem{Reck69}
R.A. Reck and D.L. Fry.
\newblock {\em Phys. Rev.}, 184:492, 1969.

\bibitem{Dalesio91}
L.R. Dalesio, A.J. Kozubal, and M.R. Kraimer.
\newblock {\em Proc. Int. Conf. Accel. and Large Exp. Physics Contr. Syst.,
  Tsukuba}, page~11, 1991.

\bibitem{Unser81}
K.~Unser.
\newblock {\em IEEE Trans. Nucl. Sci.}, NS-28:3, 1981.

\bibitem{Swartz95}
M.~Swartz {\em et al.}
\newblock {\em Nucl. Instr. Meth.}, A363:526, 1995.

\end{thebibliography}
\end{document}